# Dark Photon As an Extra U(1) Extension to the Standard Model With general Rotation in Kinetic Mixing

BERKANE Amina[1], BOUSSAHEL Mounir[1]

[1]Laboratoire Physique et Chimie des Matériaux, Département de Physique, Université de M'sila, Algérie

**Abstract**

An extension to the original dark photon model is proposed, by a generalization of matrix transformation with any angle θ of orthogonal rotation. Not only all the results presented by the dark photon model have been accurately reproduced, but new horizons have been opened by exploring new fields induced by this generalization, with new possibilities in the investigation and research of dark photon in particular and dark matter in general.

*E-mail addresses:* Amina.berkane@univ-msila.dz (A. Berkane), mounir.boussahel@univ-msila.dz (M. Boussahel)

# 1. Introduction

Since Vera Rubin and Kent Ford introduced dark matter in 1970 [1], the latter has gained great importance not only in astrophysics [2, 3] but also in cosmology and particularly in particle physics [4].

Recent observations have shown that dark matter occupies no less than 27% of the total energy density and 85% of the total mass of all the universe [5].

If the nature of dark matter has remained unknown until today. Theoretical physicists have not given up however, they have putted forward several hypotheses and models concerning the potential candidates constituting it [6]. Based on the principle that if the interactions between baryonic particles [7] are transported by gauge bosons such as photons, gauge bosons $W^{\pm}, Z$ and even the eight color gluons. Dark bosons [8] must also exist as mediators of interactions specific to dark matter such as WIMPs [9,10]. We can also cite other potential candidates for dark matter for example, sterile neutrinos [11], particles with Majorana masses, super-symmetrical particles, scalar particles [12] etc..

In this work, we are interested by a generalization to the lagrangian kinetic mixing terms proposed by [13] to an orthogonal rotation of any angle θ, in which is proposed an extension to the standard model containing dark matter. An idea of this kind is not new but it was proposed in the literature in different ways, [14] in the framework of the models of great unification (GU). We can even cite other works in a completely different framework, which is the framework of the noncommutative geometry [15] or GU SU(5)$\otimes$ U(1) [16] where the importance in this model is that the U(1) Lie group comes as an out-put of the model. But the interest of this work, concerns the mixing which is presented in an original way in dark photon model.

In this work, we will extend our study of the dark photon, within the framework of [13] to an orthogonal rotation of any angle θ, we reproduce all calculations made in [13] but in a more general way. We think that new horizons to a better understanding are opened, by enriching this work which we find very interesting by its original ideas.

This document is organized in this way: in section 2, we are interested in the term of kinetic mixing of the Lagrangian by proposing a rotation with any angle θ contained in the transformation between the two basis $A$ and $B$. Then we will be interested and always in this context of this general rotation to the spontaneous symmetry breaking, by calculating the expression of the mass of dark photon as a function of the angle θ. Then we give the expressions of the coupling constants and the interaction terms of the photon and the dark photon. In application, a simple but rather realistic example was studied in original dark photon approach. But extended to the transformation of the base $A_\mu^i$ to $B_\mu^i$ with an orthogonal rotation of any angle θ which is the essence of the present work. The coupling constants for different values of θ have been calculated, with the presentation of several diagrams of these constants as a function of the real mixing parameter c, where the particular case of θ = π / 4 has been exactly reproduced. Finally, we draw our conclusion.

## 2. Dark Photon ''Mixing Term with General Rotation''

As part of the standard model with an increasing gauge boson in general and in gauge theory, this kind of model has for Lie group:

$$SU(3)_C \otimes SU(2)_L \otimes U_1(1) \otimes U_2(1) \quad (1)$$

The boson associated to $U_1(1)$ is the photon and the $U_2(1)$ boson is usually named $Z'$ [17]. But in our case it will be associated with dark matter and that is called dark photon, where we can present the mixing term which is parameterized by the real parameter $c$, that the gauge invariance cannot fix it, on the contrary, it is dependent basis so it only makes sense by fixing the basis $A_\mu^1$, $A_\mu^2$ with a set of physical considerations:

$$\mathcal{L}_{mixing} = -2c F_{\mu\nu}^1 F^{2\,\mu\nu} \quad (2)$$

$F_{\mu\nu}^1$ and $F^{2\,\mu\nu}$ are the field intensities of $U_1(1)$ and $U_2(1)$ respectively, which allows us to write:

$$\mathcal{L}_{gauge} = -\frac{1}{4} F_{\mu\nu}^1 F^{1\,\mu\nu} - \frac{1}{4} F_{\mu\nu}^2 F^{2\,\mu\nu} - 2c F_{\mu\nu}^1 F^{2\,\mu\nu} \quad (3)$$

$$F_{\mu\nu}^r = \partial_\mu A_\nu^r - \partial_\nu A_\mu^r \quad (4)$$

It is therefore naturally justified that the fermions of the standard model can have interactions with the dark photon as mediator, which allows us to write the Lagrange interaction term as follows:

$$\mathcal{L}_{int} = g_1 j_1^\mu A_\mu^1 + g_2 j_2^\mu A_\mu^2 \quad (5)$$

Which can be written in the following matrix form:

$$\mathcal{L}_{int} = \begin{pmatrix} j_1^\mu & j_2^\mu \end{pmatrix} \begin{pmatrix} g_1 & 0 \\ 0 & g_2 \end{pmatrix} \begin{pmatrix} A_\mu^1 \\ A_\mu^2 \end{pmatrix} \quad (6)$$

Where $g_r$ is the corresponding coupling and $j_r^\mu$ ($r = 1,2$) is the fermionic current due to the presence of charge $U(1)_r$ [18]:

$$j_r^\mu = q_r^f \bar\psi \gamma^\mu \psi \quad (7)$$

through a transformation that is in fact an orthogonal rotation of any angle $\theta$ in the $A_\mu^1 - A_\mu^2$ basis followed by a scaling, we have:

$$\mathcal{L}_{gauge} = -\frac{1}{4}G^1_{\mu\nu}G^{1\ \mu\nu} - \frac{1}{4}G^2_{\mu\nu}G^{2\ \mu\nu} \tag{8}$$

Where :

$$G^r_{\mu\nu} = \partial_\mu B^r_\nu - \partial_\nu B^r_\mu \tag{9}$$

With $r = 1,2$

The main thing in all this is the disappearance of the mixing term in the new basis $B^1_\mu, B^2_\mu$ which is defined by:

$$\begin{pmatrix} A^1_\mu \\ A^2_\mu \end{pmatrix} = \frac{1}{2}\begin{pmatrix} \cos\theta\sqrt{\frac{1}{\lambda_1}} & -\sin\theta\sqrt{\frac{1}{\lambda_2}} \\ \sin\theta\sqrt{\frac{1}{\lambda_1}} & \cos\theta\sqrt{\frac{1}{\lambda_2}} \end{pmatrix}\begin{pmatrix} B^1_\mu \\ B^2_\mu \end{pmatrix} \tag{10}$$

But we lost the diagonal form of the interaction term in the new basis due to the transformation matrix given in equation (10).

When the development of (10) gives:

$$A^1_\mu = \frac{1}{2}\sqrt{\frac{1}{\lambda_1}}\cos\theta\ B^1_\mu - \frac{1}{2}\sin\theta\sqrt{\frac{1}{\lambda_2}}B^2_\mu$$

$$A^2_\mu = \frac{1}{2}\sqrt{\frac{1}{\lambda_1}}\sin\theta\ B^2_\mu + \frac{1}{2}\cos\theta\sqrt{\frac{1}{\lambda_2}}B^2_\mu$$

And by replacing in eq. (2) after a simple calculation which is not too long, one obtains the following relations of the parameters $\lambda_1$ and $\lambda_2$ successively:

$$\lambda_1 = \frac{1}{4} + 2c\cos\theta\sin\theta \tag{11}$$

and

$$\lambda_2 = \frac{1}{4} - 2c\cos\theta\sin\theta \tag{12}$$

With :

$$\lambda_1 + \lambda_2 = \frac{1}{2} \tag{13}$$

For $c > 0$ :

$$\lambda_1 > \lambda_2 \text{ for } \theta = \left]0, \frac{\pi}{2}\right[ \cup \left]\pi, \frac{3\pi}{2}\right[ \tag{14}$$

and

$$\lambda_2 > \lambda_1 \text{ for } \theta = \left]\frac{\pi}{2}, \pi\right[ \cup \left]\frac{3\pi}{2}, 2\pi\right[ \tag{15}$$

for $0 < \lambda_{1,2} < \lambda_1 + \lambda_2$  we find $c < \left|\frac{1}{8\cos\theta\sin\theta}\right|$

From (13) we have: $0 < \lambda_{1,2} < \frac{1}{2}$

By replacing with the expression of $\lambda_1$ see (11), we obtain immediately:

$$0 < \frac{1}{4} + 2c\cos\theta\sin\theta < \frac{1}{2}$$

Or:

$$0 < c < \frac{1}{8\cos\theta\sin\theta}$$

We see that for $\theta = \frac{\pi}{4}$, we get the special case :

$$0 < c < \frac{1}{4}$$

    The same thing can be done for $\lambda_2$ , to get the same result. This is actually equivalent to processing at $c \leftrightarrow -c$ to get $\lambda_1 \leftrightarrow \lambda_2$ . So we can keep $c > 0$ and working with $\lambda_1 > \lambda_2$ .

    In what follows, we will consider that the basis $B$ is orthonormal, where we will define the charges $U(1) \times U(1)$, knowing that on this basis the non-diagonal interactions with the fermions are present.

And eq. (6) becomes:

$$\mathcal{L}_{int} = \frac{1}{2}\begin{pmatrix} j_1^\mu & j_2^\mu \end{pmatrix} \begin{pmatrix} \dfrac{g_1 \cos\theta}{\sqrt{\lambda_1}} & \dfrac{-g_1 \sin\theta}{\sqrt{\lambda_2}} \\ \dfrac{g_2 \sin\theta}{\sqrt{\lambda_1}} & \dfrac{g_2 \cos\theta}{\sqrt{\lambda_2}} \end{pmatrix} \begin{pmatrix} B_\mu^1 \\ B_\mu^2 \end{pmatrix} \qquad (16)$$

The first description that we have, is one that corresponds to equation (3) and equation (6), where there is a kinetic mixing between the gauge field intensities $U(1)$ (ie $c \neq 0$ ) and the currents which only couple to the corresponding gauge bosons, i.e. $j_r^\mu$ to $A_r^\mu$ for r = 1,2.

In equation (8) and (16), we have a second description, given by a change which affects only the scaling in the matrix which is present in (16).

At the limit $c \to 0$ we have:

$$\mathcal{L}_{int} = \frac{1}{2\sqrt{2}}\begin{pmatrix} j_1^\mu & j_2^\mu \end{pmatrix} \begin{pmatrix} g_1 & -g_1 \\ g_2 & g_2 \end{pmatrix} \begin{pmatrix} B_\mu^1 \\ B_\mu^2 \end{pmatrix} \qquad (17)$$

The development of the Lagrangian interactions term in equation (16) and after of course defining the currents $J_i^\mu$. We have then:

$$\mathcal{L}_{int} = \begin{pmatrix} J_1^\mu & J_2^\mu \end{pmatrix} \begin{pmatrix} \tilde{g}_1 & 0 \\ 0 & \tilde{g}_2 \end{pmatrix} \begin{pmatrix} B_\mu^1 \\ B_\mu^2 \end{pmatrix} \qquad (18)$$

by replacing $\tilde{g}_1$ and $\tilde{g}_2$ by their expressions which are given bellow in eq.(22), we have:

$$\mathcal{L}_{int} = \begin{pmatrix} J_1^\mu & J_2^\mu \end{pmatrix} \begin{pmatrix} \dfrac{\sqrt{g_1^2 + g_2^2}}{2\sqrt{\lambda_1}} & 0 \\ 0 & \dfrac{\sqrt{g_1^2 + g_2^2}}{2\sqrt{\lambda_2}} \end{pmatrix} \begin{pmatrix} B_\mu^1 \\ B_\mu^2 \end{pmatrix} \qquad (19)$$

And by doing the same thing for $\lambda_1$ and $\lambda_2$ what will allow us to write:

$$\mathcal{L}_{int} = \begin{pmatrix} J_1^\mu & J_2^\mu \end{pmatrix} \begin{pmatrix} \dfrac{\sqrt{g_1^2+g_2^2}}{2\sqrt{\frac{1}{4}+2c\sin\theta\cos\theta}} & 0 \\ 0 & \dfrac{\sqrt{g_1^2+g_2^2}}{2\sqrt{\frac{1}{4}-2c\sin\theta\cos\theta}} \end{pmatrix} \begin{pmatrix} B_\mu^1 \\ B_\mu^2 \end{pmatrix} \quad (20)$$

In this transformation, the currents involving fermions are scaled by the following rotation:

$$\begin{pmatrix} J_1^\mu \\ J_2^\mu \end{pmatrix} = \begin{pmatrix} \cos\varphi & \sin\varphi \\ -\cos\varphi & \sin\varphi \end{pmatrix} \begin{pmatrix} j_1^\mu \\ j_2^\mu \end{pmatrix} \quad (21)$$

Which is a non-orthogonal transformation, where:

$$\cos\varphi = \frac{g_1}{\sqrt{g_1^2+g_2^2}}, \sin\varphi = \frac{g_2}{\sqrt{g_1^2+g_2^2}},$$

$$\tilde{g}_1 = \frac{\sqrt{g_1^2+g_2^2}}{2\sqrt{\lambda_1}}, \tilde{g}_2 = \frac{\sqrt{g_1^2+g_2^2}}{2\sqrt{\lambda_2}} \quad (22)$$

For $c > 0$ with $\lambda_1 > \lambda_2$ and leads $\tilde{g}_2 > \tilde{g}_1$

In the case $g_1 = g_2 = g$, the relations in equation (17) become:

$$\cos\varphi = \sin\varphi = \frac{1}{\sqrt{2}}, \tilde{g}_1 = \frac{g}{2\sqrt{\lambda_1}}, \tilde{g}_2 = \frac{g}{2\sqrt{\lambda_2}}$$

$$\cos\varphi = \sin\varphi = \frac{1}{\sqrt{2}}, \tilde{g}_1 = \frac{\sqrt{g_1^2+g_2^2}}{2\sqrt{\frac{1}{4}+2c\sin\theta\cos\theta}}, \tilde{g}_2 = \frac{\sqrt{g_1^2+g_2^2}}{2\sqrt{\frac{1}{4}-2c\sin\theta\cos\theta}} \quad (23)$$

## 3. Extension of generalized rotation to $U^1(1) \times U^2(1)$: **spontaneous symmetry breaking:**

In this section we will extend the generalization of rotation to spontaneous symmetry breaking (SSB). For a Scalar field $\Phi$ with $U^{1,2}(1)$ charges $q_{1,2}^S$ loads the covariant derivative:

$$D^\mu \varphi = [\partial^\mu - ig_1 q_1^s A_1^\mu - ig_2 q_2^s A_2^\mu]\Phi$$

$$= [\partial^\mu - i\widetilde{g_1} Q_1^s B_1^\mu - i\tilde{g}_2 Q_2^s B_2^\mu]\Phi$$

$$= \left[\partial^\mu - i\frac{\sqrt{g_1^2 + g_2^2}}{2\sqrt{\frac{1}{4} + 2c\sin\theta\cos\theta}} Q_1^s B_1^\mu - i\frac{\sqrt{g_1^2 + g_2^2}}{2\sqrt{\frac{1}{4} - 2c\sin\theta\cos\theta}} Q_2^s B_2^\mu\right]\Phi \quad (24)$$

The charges $Q_i$ are defined in the basis $B$ and the charges $q_i$ are in the basis $A$, they are related to each other by the same way as equation (21).

Thus :

$$\begin{pmatrix} Q_1 \\ Q_2 \end{pmatrix} = \begin{pmatrix} \cos\varphi & \sin\varphi \\ -\cos\varphi & \sin\varphi \end{pmatrix} \begin{pmatrix} q_1 \\ q_2 \end{pmatrix} \quad (25)$$

For the case of the vev : $<\Phi> = \frac{v}{\sqrt{2}} \neq 0$ [19] . The mass matrix of the gauge boson in the basis $B_{1,2}$ is given by:

$$M_{gauge}^2 = \frac{v^2}{2}\begin{pmatrix} (\tilde{g}_1 Q_1^s)^2 & (\tilde{g}_2 Q_2^s)(\tilde{g}_1 Q_1^s) \\ (\tilde{g}_2 Q_2^s)(\tilde{g}_1 Q_1^s) & (\tilde{g}_2 Q_2^s)^2 \end{pmatrix}$$

By replacing the $\tilde{g}_i$ by their values we obtain:

$$M_{gauge}^2 = \frac{v^2}{2}\begin{pmatrix} \left(\frac{\sqrt{g_1^2 + g_2^2}}{2\sqrt{\frac{1}{4} + 2c\sin\theta\cos\theta}} Q_1^s\right)^2 & \left(\frac{\sqrt{g_1^2 + g_2^2}}{2\sqrt{\frac{1}{4} - 2c\sin\theta\cos\theta}} Q_2^s\right)\left(\frac{\sqrt{g_1^2 + g_2^2}}{2\sqrt{\frac{1}{4} + 2c\sin\theta\cos\theta}} Q_1^s\right) \\ \left(\frac{\sqrt{g_1^2 + g_2^2}}{2\sqrt{\frac{1}{4} - 2c\sin\theta\cos\theta}} Q_2^s\right)\left(\frac{\sqrt{g_1^2 + g_2^2}}{2\sqrt{\frac{1}{4} + 2c\sin\theta\cos\theta}} Q_1^s\right) & \left(\frac{\sqrt{g_1^2 + g_2^2}}{2\sqrt{\frac{1}{4} - 2c\sin\theta\cos\theta}} Q_2^s\right)^2 \end{pmatrix} \quad (26)$$

Let the orthogonal fundamental mass states $X_\mu^1$, $X_\mu^2$ which are eigenvectors of a real symmetrical matrix with distinct eigenvalues:

$$\begin{pmatrix} X_\mu^1 \\ X_\mu^2 \end{pmatrix} = \begin{pmatrix} \cos\chi & -\sin\chi \\ \sin\chi & \cos\chi \end{pmatrix} \begin{pmatrix} B_\mu^1 \\ B_\mu^2 \end{pmatrix} \quad (27)$$

The mixing angle $\chi$ is defined by:

$$\cos\chi = \frac{1}{N}|\tilde{g}_2 Q_2^s| \quad , \quad \sin\chi = \frac{1}{N}|\tilde{g}_1 Q_1^s|$$

with more details we have:

$$\cos\chi = \frac{1}{N}\left|\frac{\sqrt{g_1^2+g_2^2}}{2\sqrt{\frac{1}{4}-2c\sin\theta\cos\theta}}Q_2^s\right|, \quad \sin\chi = \frac{1}{N}\left|\frac{\sqrt{g_1^2+g_2^2}}{2\sqrt{\frac{1}{4}+2c\sin\theta\cos\theta}}Q_1^s\right| \quad (28)$$

the normalization factor is given by $N$:

$$N^2 = (\tilde{g}_2 Q_2^s)^2 + (\tilde{g}_1 Q_1^s)^2$$

in the same way as equation (25) we have:

$$N^2 = \left(\frac{\sqrt{g_1^2+g_2^2}}{2\sqrt{\frac{1}{4}-2c\sin\theta\cos\theta}}Q_2^s\right)^2 + \left(\frac{\sqrt{g_1^2+g_2^2}}{2\sqrt{\frac{1}{4}+2c\sin\theta\cos\theta}}Q_1^s\right)^2 \quad (29)$$

There are two eigenvalues of the mass matrix which are:

$$m_1^2 = 0 \ , m_2^2 = N^2 v^2$$

$$m_1^2 = 0, m_2^2 = \left[\left(\frac{\sqrt{g_1^2+g_2^2}}{2\sqrt{\frac{1}{4}-2c\sin\theta\cos\theta}}Q_2^s\right)^2 + \left(\frac{\sqrt{g_1^2+g_2^2}}{2\sqrt{\frac{1}{4}+2c\sin\theta\cos\theta}}Q_1^s\right)^2\right]v^2 \quad (30)$$

This allows us to write from equation (20) the interactions term of the mass eigen states $X_\mu^1$ and $X_\mu^2$ as follows:

$$\mathcal{L}_{int} = \begin{pmatrix} J_1^\mu & J_2^\mu \end{pmatrix} \begin{pmatrix} \tilde{g}_1 & 0 \\ 0 & \tilde{g}_2 \end{pmatrix} \begin{pmatrix} \cos\chi & \sin\chi \\ -\sin\chi & \cos\chi \end{pmatrix} \begin{pmatrix} X_\mu^1 \\ X_\mu^2 \end{pmatrix}$$

by replacing $\tilde{g}_1$ and $\tilde{g}_2$ by their expressions we have:

$$\mathcal{L}_{int} = \begin{pmatrix} J_1^\mu & J_2^\mu \end{pmatrix} \begin{pmatrix} \frac{\sqrt{g_1^2+g_2^2}}{2\sqrt{\frac{1}{4}+2c\sin\theta\cos\theta}} & 0 \\ 0 & \frac{\sqrt{g_1^2+g_2^2}}{2\sqrt{\frac{1}{4}-2c\sin\theta\cos\theta}} \end{pmatrix} \begin{pmatrix} \cos\chi & \sin\chi \\ -\sin\chi & \cos\chi \end{pmatrix} \begin{pmatrix} X_\mu^1 \\ X_\mu^2 \end{pmatrix} \quad (31)$$

After the $U(1) \times U(1)$ spontaneous symmetry breaking, the result is an associated conserved charge, which is a linear combination of $Q_1$ and $Q_2$. This conserved charge can be written in a normalized form:

$$Q = \alpha_1 Q_1 + \alpha_2 Q_2 \quad , \quad \alpha_1^2 + \alpha_2^2 = 1 \tag{32}$$

While the standard field, which acquires the vev, must satisfy:

$$\alpha_1 Q_1^s + \alpha_2 Q_2^s = 0 \tag{33}$$

Which implies:

$$\alpha_1 = \frac{Q_2^s}{\sqrt{Q_1^{s\,2} + Q_2^{s\,2}}}, \quad \alpha_2 = \frac{-Q_1^s}{\sqrt{Q_1^{s\,2} + Q_2^{s\,2}}} \tag{34}$$

Another non conserved charge $Q'$ can also be defined, which is orthogonal to $Q$ such as:

$$Q' = -\alpha_2 Q_1 + \alpha_1 Q_2 \tag{35}$$

$Q'$ is not conserved because the $U(1)$ symmetry is broken. From equation (34) we can calculate the mass of $X_\mu^2$ in terms of $\alpha_{1,2}$ :

$$m_2^2 = \frac{(g_1^2 + g_2^2)(Q_1^{s\,2} + Q_2^{s\,2})}{8}\left(\frac{\alpha_1^2}{\lambda_2} + \frac{\alpha_2^2}{\lambda_1}\right)$$

$$m_2^2 = \frac{(g_1^2 + g_2^2)(Q_1^{s\,2} + Q_2^{s\,2})}{8}\left(\frac{\alpha_1^2}{\frac{1}{4} - 2c\sin\theta\cos\theta} + \frac{\alpha_2^2}{\frac{1}{4} + 2c\sin\theta\cos\theta}\right)v^2 \tag{36}$$

## 4. Effects of rotation generalization extended to Fermion interactions:

We can express the fermionic interactions of mass eigenstates of the gauge bosons in terms of currents defined by the charges $Q^f$ and $Q'$, taking in to account all the changes introduced in the previous sections, where we have the task above all to show everything that will appear to be different compared to the case of the orthogonal angle rotation $\frac{\pi}{4}$:

$$\widehat{J_1^\mu} = Q^f \overline{\psi}\gamma^\mu \psi = (J_1^\mu \alpha_1 + J_2^\mu \alpha_2) \quad , \quad \widehat{J_2^\mu} = Q'^f \overline{\psi}\gamma^\mu \psi = -J_1^\mu \alpha_2 + J_2^\mu \alpha_1 \tag{37}$$

The Lagrangian interactions term for massive and massless gauge bosons is written as:

$$\mathcal{L}_{int} = \sum_{i,j=1,2} g_{ij} J_i^\mu X_\mu^j \tag{38}$$

$X_\mu^1$ corresponds to the residual $U(1)$ and it couples only to $\widehat{J_1^\mu}$, but $X_\mu^2$ couple once to $\widehat{J_1^\mu}$ and to the orthogonal combination, more precisely $\widehat{J_2^\mu}$. To determine the coupling constants $g_{ij}$, we rewrite equation (31) by:

$$\mathcal{L}_{int} = [(\tilde{g}_1 J_1^\mu \cos\chi - \tilde{g}_2 J_2^\mu \sin\chi)X_\mu^1 + (\tilde{g}_1 J_1^\mu \sin\chi + \tilde{g}_2 J_2^\mu \cos\chi)X_\mu^2]$$

$$= \left[\left(\frac{\sqrt{g_1^2+g_2^2}}{2\sqrt{\frac{1}{4}+2c\sin\theta\cos\theta}} J_1^\mu \cos\chi - \frac{\sqrt{g_1^2+g_2^2}}{2\sqrt{\frac{1}{4}-2c\sin\theta\cos\theta}} J_2^\mu \sin\chi\right)X_\mu^1 \right.$$

$$\left. + \left(\frac{\sqrt{g_1^2+g_2^2}}{2\sqrt{\frac{1}{4}+2c\sin\theta\cos\theta}} J_1^\mu \sin\chi + \frac{\sqrt{g_1^2+g_2^2}}{2\sqrt{\frac{1}{4}-2c\sin\theta\cos\theta}} J_2^\mu \cos\chi\right)X_\mu^2\right] \tag{39}$$

Using equations (28) and (34), the term interactions of the massless gauge boson $X_\mu^1$:

$$\mathcal{L}_{X^1} = \frac{\tilde{g}_1 \tilde{g}_2}{\sqrt{\tilde{g}_1^2 \alpha_2^2 + \tilde{g}_2^2 \alpha_1^2}} (J_1^\mu \alpha_1 + J_2^\mu \alpha_2) X_\mu^1 = g_{11} J_1^\mu X_1^\mu$$

Where :

$$\mathcal{L}_{X^1} = \sqrt{\frac{g_1^2 + g_2^2}{\frac{1}{8}\left[\left(\frac{1}{4}+2c\sin\theta\cos\theta\right)\alpha_2^2 + \left(\frac{1}{4}-2c\sin\theta\cos\theta\right)\alpha_1^2\right]}} (J_1^\mu \alpha_1 + J_2^\mu \alpha_2) X_\mu^1 \tag{40}$$

$$= g_{11} J_1^\mu X_1^\mu$$

We can get the coupling constants $g_{11}$ and $g_{21}$ from equation (40):

$$g_{11} = \frac{\tilde{g}_1 \tilde{g}_2}{\sqrt{\tilde{g}_1^2 \alpha_2^2 + \tilde{g}_2^2 \alpha_1^2}} \quad , g_{21} = 0$$

$$g_{11} = \sqrt{g_1^2 + g_2^2} \frac{2\sqrt{2}}{\sqrt{\left(\frac{1}{4}+2c\sin\theta\cos\theta\right)\alpha_2^2+\left(\frac{1}{4}-2c\sin\theta\cos\theta\right)\alpha_1^2}} \quad , \quad g_{21} = 0 \tag{41}$$

Equation (40) gives the familiar expression:

$$\frac{1}{g_{11}^2} = \frac{\alpha_1^2}{\tilde{g}_1^2} + \frac{\alpha_2^2}{\tilde{g}_2^2} \tag{42}$$

As a result of equation (42) we find:

$$\tilde{g}_1 \leq g_{11} \leq \tilde{g}_2$$

$$\frac{\sqrt{g_1^2 + g_2^2}}{2\sqrt{\frac{1}{4} + 2c\sin\theta\cos\theta}} \leq g_{11} \leq \frac{\sqrt{g_1^2 + g_2^2}}{2\sqrt{\frac{1}{4} - 2c\sin\theta\cos\theta}} \tag{43}$$

$X_\mu^1$ Corresponds to a symmetry $U(1)$, it is coupled only to $\widehat{J_1^\mu}$. The interactions of $X_\mu^2$ can be expressed as:

$$\mathcal{L}_{X^2} = \frac{1}{\sqrt{\tilde{g}_1^2 \alpha_2^2 + \tilde{g}_2^2 \alpha_1^2}} \left( -\alpha_1 \alpha_2 (\tilde{g}_1^2 - \tilde{g}_2^2) \hat{J}_1^\mu + (\alpha_2^2 \tilde{g}_1^2 + \alpha_1^2 \tilde{g}_2^2) \hat{J}_2^\mu \right) X_\mu^2$$

$$\begin{aligned}
\mathcal{L}_{X^2} = \frac{\sqrt{g_1^2 + g_2^2}}{\sqrt{\frac{\alpha_2^2}{\frac{1}{4}+2c\sin\theta\cos\theta} + \frac{\alpha_1^2}{\frac{1}{4}-2c\sin\theta\cos\theta}}} &\left( -\alpha_1\alpha_2 \left( \frac{1}{\frac{1}{4}+2c\sin\theta\cos\theta} \right. \right. \\
&\left. - \frac{1}{\frac{1}{4}-2c\sin\theta\cos\theta} \right) \hat{J}_1^\mu \\
&\left. + \left( \frac{\alpha_2^2}{\frac{1}{4}+2c\sin\theta\cos\theta} + \frac{\alpha_1^2}{\frac{1}{4}-2c\sin\theta\cos\theta} \right) \hat{J}_2^\mu \right) X_\mu^2
\end{aligned} \tag{44}$$

From there, we can make a new reading of the couplings of the massif boson $X_\mu^2$, with the two currents, namely:

$$g_{22} = \sqrt{\tilde{g}_1^2 \alpha_2^2 + \tilde{g}_2^2 \alpha_1^2} = \frac{\sqrt{g_1^2 + g_2^2}}{2\sqrt{2}} \sqrt{\left( \frac{\alpha_2^2}{\lambda_1} + \frac{\alpha_1^2}{\lambda_2} \right)}$$

$$= \frac{\sqrt{g_1^2 + g_2^2}}{2\sqrt{2}} \sqrt{\left(\frac{\alpha_2^2}{\frac{1}{4} + 2c\sin\theta\cos\theta} + \frac{\alpha_1^2}{\frac{1}{4} - 2c\sin\theta\cos\theta}\right)} \quad (45)$$

and :

$$g_{12} = \frac{\alpha_1 \alpha_2 (\tilde{g}_1^2 - \tilde{g}_2^2)}{\sqrt{\tilde{g}_1^2 \alpha_2^2 + \tilde{g}_2^2 \alpha_1^2}} = -\alpha_1 \alpha_2 \frac{\sqrt{g_1^2 + g_2^2}}{2\sqrt{2}} \left(\frac{\lambda_2 - \lambda_1}{\sqrt{(\lambda_1 \lambda_2)(\alpha_1^2 \lambda_2 + \alpha_2^2 \lambda_2)}}\right)$$

$$= -\alpha_1 \alpha_2 \frac{\sqrt{g_1^2 + g_2^2}}{2\sqrt{2}} \left(\frac{4c \sin\theta \cos\theta}{\sqrt{(16 - 4c^2 \sin^2\theta \cos^2\theta)\left(\alpha_1^2 \left(\frac{1}{4} + 2c\sin\theta\cos\theta\right) + \alpha_2^2 \left(\frac{1}{4} - 2c\sin\theta\cos\theta\right)\right)}}\right) \quad (46)$$

It should be noted that nothing forbids $X_\mu^2$ to couple either to $\widehat{J}_1^\mu$ and not necessarily only to $\widehat{J}_2^\mu$.

### 5. Calculation of couplings and charges with an example of application:

The dark photon or "dark electromagnetism" was introduced in 2008 by physicists Lotty Ackerman, Matthew R. Buckley, Sean M. Carroll, and Marc Kamionkowski. The idea of the dark photon is based on the analogy with the photon of the Standard Model [20]. Indeed, the different particles that compose ordinary matter interact with each other through electromagnetic interaction, which is mediated by the photons[21].

Therefore, the hypothesis that dark matter particles could also interact with each other via bosons was proposed. Physicists postulated the existence of dark photons, present in the Hidden Sector (part of the quantum fields and unknown particles of the Standard Model) and responsible for the interactions between dark matter particles by a derived force from ordinary electromagnetism [6], buptized as dark electromagnetism.

In the same way as in the Standard Model, dark electromagnetism is based on the existence of a quantum field of gauge associated to the gauge group U(1) (mathematical group integrating the gauge symmetry and allowing to construct the electromagnetic interaction). This new force is carried by the dark photon, noted «$A'$» [22], a spin-1 boson interacting directly with dark matter and indirectly with particles of the Standard Model.

Since it is massive and Unlike the classic photon, the dark photon is a boson which must be extremely unstable precisely because of its mass which is non-null.

This instability should lead to a rapid disintegration into particles of undetectable dark matter, as it can normally disintegrate into a particle-antiparticle pair of electron-positron for example.

In the following the massless gauge boson $X_\mu^1$ will naturally be associated with the 'ordinary' photon and $g_{11}$ will be associated to the fine structure constant by:

$$g_{11} = e = \sqrt{4\pi\alpha_{EM}} \tag{47}$$

This allows us to rewrite equation (41) by the following relation:

$$\alpha_1^2 \left(\frac{e^2}{\tilde{g}_1^2}\right) + \alpha_2^2 \left(\frac{e^2}{\tilde{g}_1^2}\right) = 1 \tag{48}$$

Using (23), (11) and (12) we obtain the following equation:

$$(g_1^2 + g_2^2) = 2e^2[1 + 8c \sin\theta \cos\theta \, (\alpha_1^2 - \alpha_2^2)] \tag{49}$$

Equation (42) is used to express the couplings $\tilde{g}_1$ and $\tilde{g}_2$ with units as $e$:

$$\tilde{g}_i/e = \frac{\sqrt{g_1^2 + g_2^2}}{2\sqrt{2}e} \frac{1}{\sqrt{\lambda_1}} = \sqrt{\frac{\xi}{\lambda_i}} = \sqrt{\frac{\xi}{\frac{1}{4} + 2c \sin\theta \cos\theta}} \quad (i = 1,2) \tag{50}$$

The new parameter $\xi$ is given by:

$$\xi = \frac{1}{8}\left(\frac{g_1^2 + g_2^2}{e^2}\right) \tag{51}$$

Depending on $\xi$, equation (43) takes the following form:

$$\lambda_1 \alpha_1^2 + \lambda_2 \alpha_2^2 = \xi \tag{52}$$

$\lambda_{1,2}$ are determined by the kinetic mixing parameter $c$ using the normalized property $\alpha_1^2 + \alpha_2^2 = 1$ to obtain $\alpha_{1,2}$:

$$\alpha_1 = \sqrt{\frac{\xi - \left(\frac{1}{4} - 2c \sin\theta \cos\theta\right)}{4c \sin\theta \cos\theta}} \quad ; \quad \alpha_2 = \sqrt{\frac{\left(\frac{1}{4} - 2c \sin\theta \cos\theta\right) - \xi}{4c \sin\theta \cos\theta}} \tag{53}$$

When $0 \leq \alpha_1^2 \leq 1$ and $0 \leq \alpha_2^2 \leq 1$ we get :

$$|c| \geq \left| \frac{\xi - \frac{1}{4}}{2 \sin \theta \cos \theta} \right| \qquad (54)$$

When $\xi$ takes the value $1/4$ equation (51) becomes:

$$(g_1^2 + g_2^2) = 2e^2$$

From equations (45), (46) the two couplings $g_{12}$ and $g_{22}$ will have the following relations:

$$g_{12} = -e\alpha_1\alpha_2 \left( \sqrt{\frac{\lambda_2}{\lambda_1}} - \sqrt{\frac{\lambda_1}{\lambda_2}} \right) \quad , \quad g_{22} = e \frac{\xi}{\sqrt{\lambda_1\lambda_2}} \qquad (55)$$

Where the product $\lambda_1\lambda_2$ is given by:

$$\lambda_1\lambda_2 = \left( \frac{1}{16} - 4c^2 \sin \theta \cos \theta \right) \qquad (56)$$

What allows us to have $g_{12}$ and $g_{22}$:

$$g_{12} = -e \sqrt{\frac{\left( \xi - \left( \frac{1}{4} + 2c \sin \theta \cos \theta \right) \right) \left( \left( \frac{1}{4} - 2c \sin \theta \cos \theta \right) - \xi \right)}{\left( \frac{1}{4} + 2c \sin \theta \cos \theta \right) \left( \frac{1}{4} - 2c \sin \theta \cos \theta \right)}} \qquad (57)$$

$$g_{22} = \frac{e\,\xi}{\sqrt{\left( \frac{1}{4} + 2c \sin \theta \cos \theta \right) \left( \frac{1}{4} - 2c \sin \theta \cos \theta \right)}} \qquad (58)$$

The equations (57) and (58) will allow us to make a plot where we estimate the possible values of $g_{12}$ and $g_{22}$ see Fig (1-a,b,c,d,e,f,g ) according to the kinetic mixing parameter $c$, by a choice of different values of the rotation angle $\theta$ and only for certain values of $\xi$:

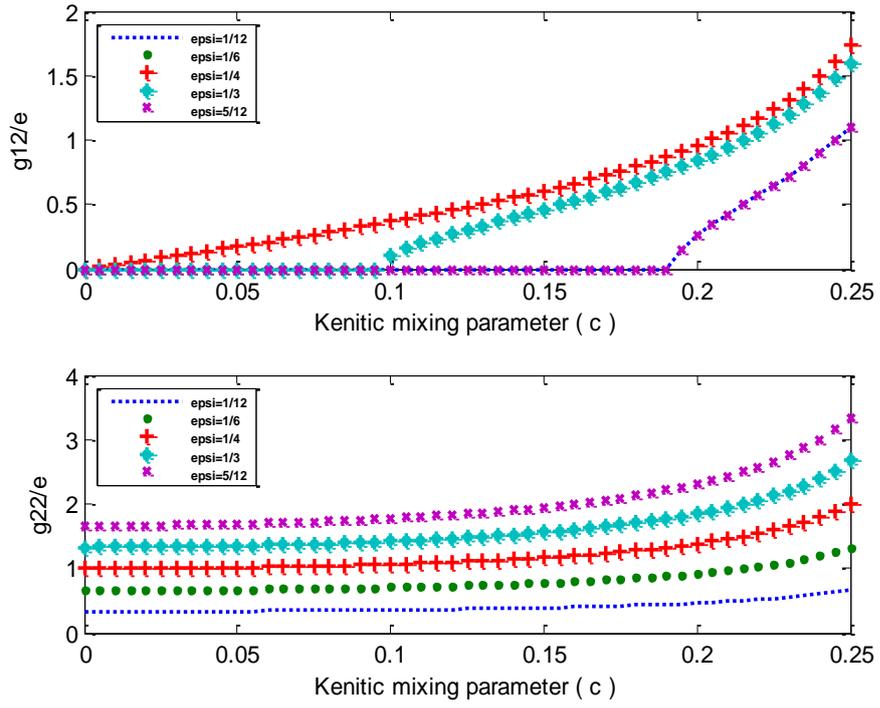

Figure 1-a: coupling $g_{12}$ and $g_{22}$ of dark photon $X_\mu^2$ according to $c$ when $g_{11} = e$ and $g_{21} = 0$ . with values different from $\xi$ and $\theta = \frac{\pi}{3}$

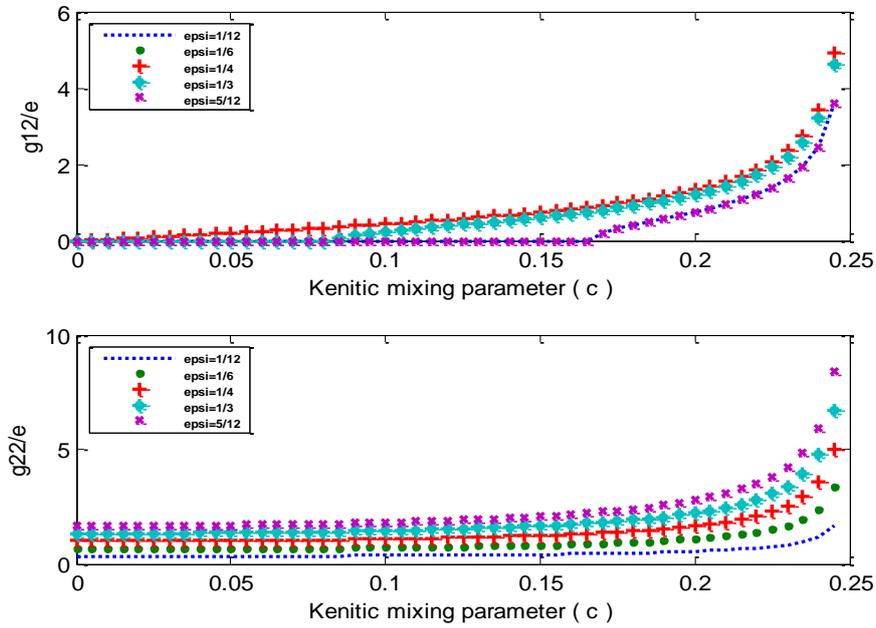

Figure 1-b : coupling $g_{12}$ and $g_{22}$ of dark photon $X_\mu^2$ according to $c$ when $g_{11} = e$ and $g_{21} = 0$ . with values different from $\xi$ and $\theta = \frac{\pi}{4}$

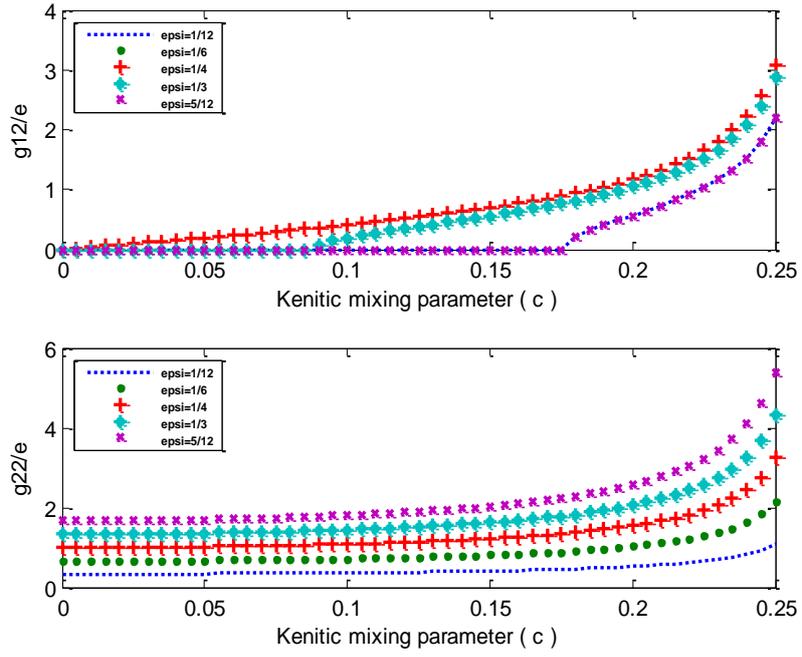

Figure 1- c : coupling $g_{12}$ and $g_{22}$ of dark photon $X_\mu^2$ according to $c$ when $g_{11} = e$ and $g_{21} = 0$ . with values different from $\xi$ and $\theta = \frac{\pi}{5}$

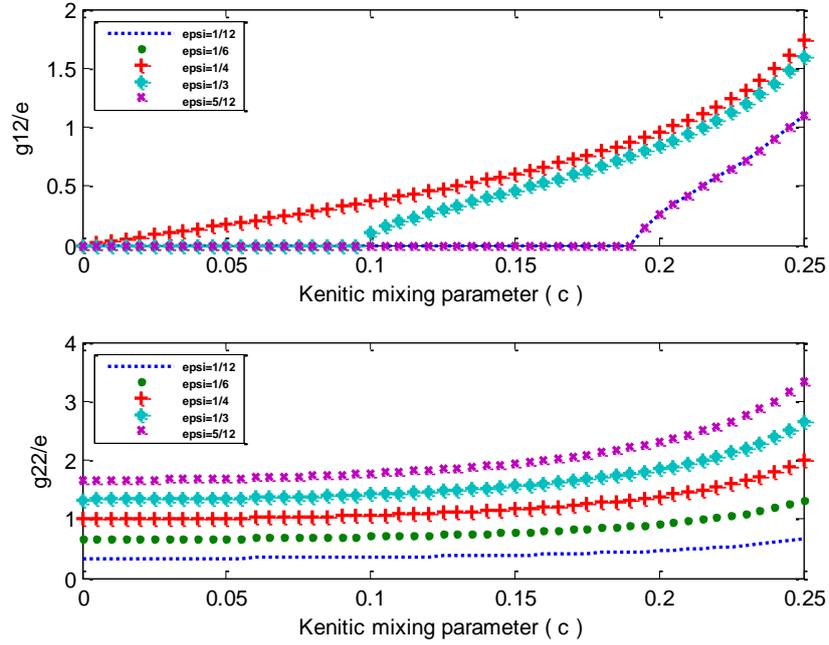

Figure 1- d : coupling $g_{12}$ and $g_{22}$ of dark photon $X_\mu^2$ according to $c$ when $g_{11} = e$ and $g_{21} = 0$ . with values different from $\xi$ and $\theta = \frac{\pi}{6}$

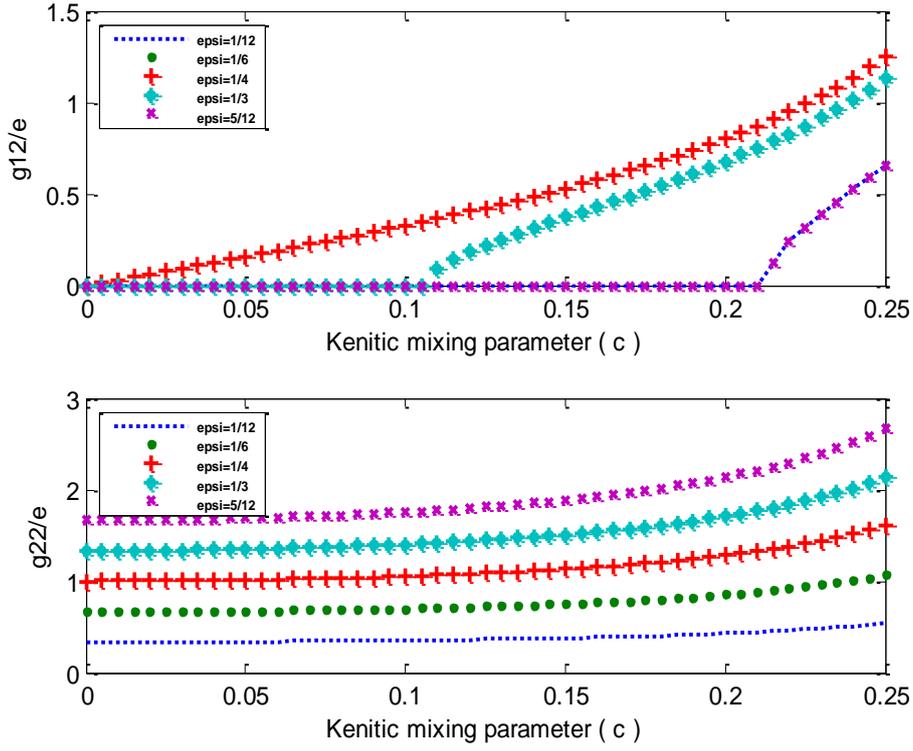

Figure 1- e : coupling $g_{12}$ and $g_{22}$ of dark photon $X_\mu^2$ according to $c$ when $g_{11} = e$ and $g_{21} = 0$ . with values different from $\xi$ and $\theta = \frac{\pi}{7}$

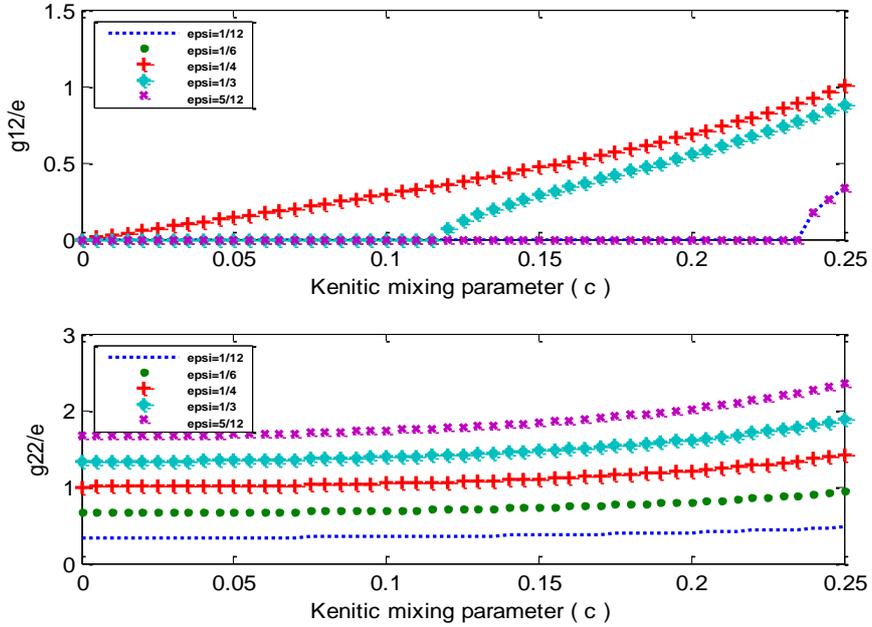

Figure 1- f : coupling $g_{12}$ and $g_{22}$ of dark photon $X_\mu^2$ according to $c$ when $g_{11} = e$ and $g_{21} = 0$ . with values different from $\xi$ and $\theta = \frac{\pi}{8}$

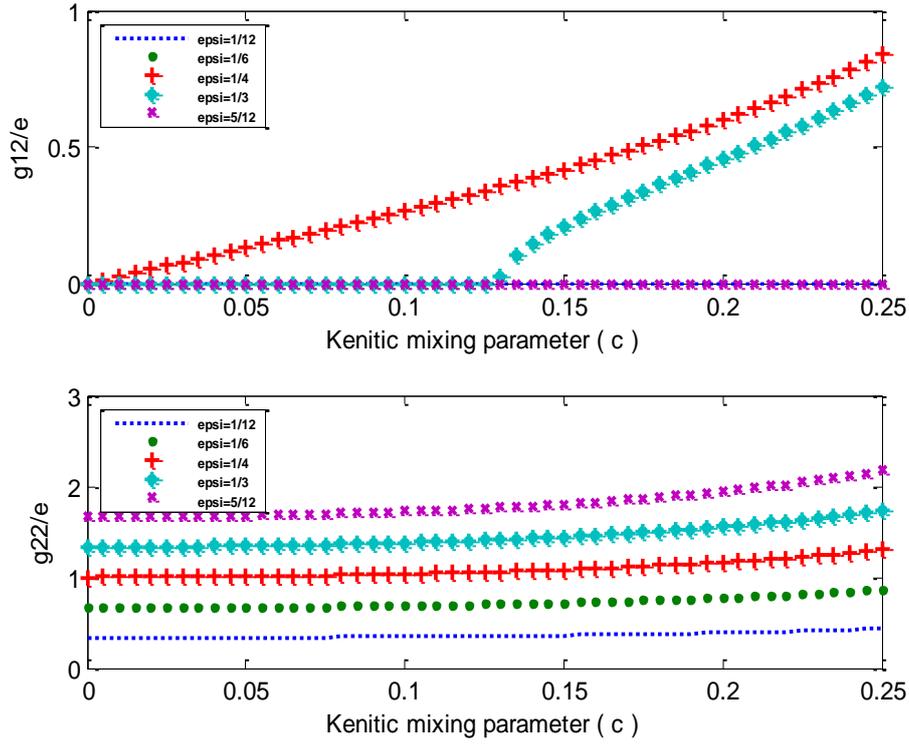

Figure 1 - g: coupling $g_{12}$ and $g_{22}$ of dark photon $X_\mu^2$ according to $c$ when $g_{11} = e$ and $g_{21} = 0$. with values different from $\xi$ and $\theta = \frac{\pi}{9}$

## 6. Discussion and conclusion:

In both diagrams, curves are displayed that correspond to several and different choices. We choose the values of $\xi$ across a range of allowed values. We explain in Fig.1 the behavior of $g_{22}$ and $g_{12}$. We take the average value of $\xi = \frac{1}{4}$ and other values above or below this average value.

For values of $\xi = \frac{1}{12}, \frac{1}{6}, \frac{1}{4}, \frac{1}{3}$ and $\theta = \frac{\pi}{3}, \frac{\pi}{4}, \frac{\pi}{5}, \frac{\pi}{6}, \frac{\pi}{7}, \frac{\pi}{8}, \frac{\pi}{9}$ It is observed that for $\theta = \frac{\pi}{3}$ and $c \in [0; 0.1]$ : $g_{12}$ takes a non-zero value for $\xi = \frac{1}{4}$ but it eliminate itself when $c \in [0.1; 0.15]$ for $\xi = \frac{1}{12}, \frac{1}{6}$, finally $g_{12}$ increases for

$c \geq 0.15$ and $\xi = \frac{1}{12}, \frac{1}{6}$.

with case for $\xi = \frac{1}{3}$: $g_{12}$ is null when $c \in [0; 0.15]$ and it increases for $c \geq 0.15$ but $g_{22}$ increases when $c \in [0; 0.25]$ and for all values of $\xi$.

The same observation is for the other angles $\theta$ with little change in value of $g_{12}$ since $\xi = \frac{1}{3}, \frac{1}{12}, \frac{6}{12}$. but for lower values of $c$, $g_{12}$ is null while $g_{22}$ takes a certain value.

The same explanation applies to the mass $m_2$ of the dark photon.

Using equations (36) and (52) we find:

$$m_2^2 = \frac{e^2 \xi^2 (Q_1^2 + Q_2^2)}{\left(\frac{1}{16} - 4c^2 \sin^2\theta \cos^2\theta\right)} v^2 \qquad (59)$$

For $c$ negative and exchanging $\alpha_1$ with $\alpha_2$, while $\lambda_1$ and $\lambda_2$ are also exchanged. Which will turn $g_{12}$ into $-g_{21}$ then that $g_{22}$ will not be affected.

As an example of application as a simple and realistic model that often appears in the kinetic mixing literature, in which the photon mixes with a gauge field $U(1)_{dm}$ because the other gauge boson is not yet experimentally detected, the symmetry of $U(1)_{dm}$ is broken and the dark photon is of very large mass. Typically, the mixing term is considered as a perturbation and its effects are examined. However, in the "Dark photon" approach, which we have extended to a generalized rotation, the remaining broken symmetry must be identified as QED, which is also one of the two initial $U(1)$ symmetries. Thus, it is therefore necessary to require that $X_\mu^1 \equiv A_1^\mu$. From equations (25) and (32), we can write:

$$Q = q^1 (\alpha_1 - \alpha_2) \cos\varphi + q^2 (\alpha_1 + \alpha_2) \sin\varphi \qquad (60)$$

With the condition that $Q = q^1$ can be reached when:

$$\left\{\alpha_1 = -\alpha_2 = \frac{1}{\sqrt{2}}, \varphi = \frac{\pi}{4}\right\}, \{\varphi = 0, \alpha_2 = 0, \alpha_1 = 1\} \qquad (61)$$

Among these, the second option is untenable because it implies that $g_2 = 0$, is in fact a consequence of equation (23). This result identifies the electrical charge as the coupling of one of the groups of factors that existed before the symmetry breaking.

$$g_1 = g_2 = g \quad , \tilde{g}_1 = \frac{g}{2\sqrt{\lambda_1}} \quad , \tilde{g}_2 = \frac{g}{2\sqrt{\lambda_2}} \qquad (62)$$

Thus, in the $A$ basis where the gauge bosons have diagonal couplings with fermions, the gauge coupling must be identical for both $U(1)$ factors. We confirm the newest of these results and we completely agree to its originality with [13].

Then equations (55), (57) and (58),

For : $\xi \to \frac{1}{4}$

$$g_{11} = e \tag{63}$$

$$g_{22} = \frac{e}{\sqrt{1 - 64c\sin^2\theta\cos^2\theta^2}} \tag{64}$$

$$g_{12} = -e\sqrt{\frac{4c^2\sin\theta^2\cos\theta^2}{\left(\frac{1}{16} - 4c^2\sin\theta^2\cos\theta^2\right)}} \tag{65}$$

Using equation (51), we obtain $\xi = \frac{1}{4}$ for which, as mentioned above $c \geq 0$.

Before finishing, we can mention the Stückelberg mechanism. Where the mixing is defined in a basis in which the gauge bosons are already at proper mass states, one of them being massless, the other having a non-zero mass [23]. In such scenarios, the elimination of the kinetic mixing is made possible by the transformation

for $\xi \to \frac{1}{4}$:

$$\begin{pmatrix} A_\mu^1 \\ A_\mu^2 \end{pmatrix} = \begin{pmatrix} 1 & -e\sqrt{\frac{4c^2\sin\theta^2\cos\theta^2}{\left(\frac{1}{16} - 4c^2\sin\theta^2\cos\theta^2\right)}} \\ 0 & \frac{e}{\sqrt{1 - 64c\sin^2\theta\cos^2\theta^2}} \end{pmatrix} \begin{pmatrix} X_\mu^1 \\ X_\mu^2 \end{pmatrix} \tag{66}$$

Now let us to conclude. In fact, from a theoretical point of view, there is nothing to prevent the generalization of well-determined orthogonal rotation to any rotation with any $\theta$, on the contrary, if it is a good generalisation, it may open up other horizons for research, especially on an experimental or observational level. In some model, when you have a transformation to use that is fundamental to the future of the work, which is in fact the case of the work proposed by [13]. At the beginning, it is good to start with a simple and particular case to better see, understand and to develop a clear image of the contour of the problem from a theoretical, phenomenological or experiential point of view. But sooner or later generalization will be necessary by itself, to expect and deal with any question and may be a possible new observation or detection of experimental type. It is for this reason and in order to prevent these kinds of questions that we have proposed this document, which is only a modest contribution and a simple generalization but which we consider useful to the original work proposed by [13]. It turned out that this generalization was able to touch all the interesting parts already seen by the original document. In the case of an extension to the standard model by Group $U(1)$ in addition, this makes it possible to have kinetic mixing terms in a more general frame work induced by the transformation of the $A_\mu$ to $B_\mu$ basis by a rotation of any angle $\theta$.

After having extended this generalization to the mixing term of the Lagrangian and with consideration of the dark photon all the formulation of the old model was generalized and studied. Then our study was extended to the spontaneous symmetry breaking, which also means to the masses of the gauge bosons of course which were also reformulated see eq (26), (30) and (36).

The interest in the terms of interactions, mainly allowed us to define the parameters of the dark photon with the fermionic current parameters $g_{12}$ and $g_{22}$. In application of the realistic and simple example, we made a plot of $g_{12}$ and $g_{22}$ see Fig 1(a,b,c,d,e,f) proposed thanks to various choices of the mixing parameter $c$ and angle $\theta$.

Finally, $\theta$ can be fixed only by knowing the exact value of the mass of the dark photon.

**Acknowledgements**

We are very grateful to the Algerian Minister of Higher Education and Scientific Research and DGRSDT for the financial support.